\documentclass[apj]{emulateapj} 
\usepackage[colorlinks=true,urlcolor=blue,citecolor=blue,linkcolor=blue]{hyperref}
\usepackage{twoopt}

\bibpunct{(}{)}{;}{a}{}{,}    
\makeatletter
 \newcommandtwoopt{\citeads}[3][][]{%
   \nonstopmode
   \href{http://adsabs.harvard.edu/abs/#3}%
        {\def\hyper@linkstart##1##2{}%
         \let\hyper@linkend\@empty\citealp[#1][#2]{#3}}
   \biblink{#3}{\href{http://adsabs.harvard.edu/abs/#3}{ADS}}%
   \errorstopmode}            
 \newcommandtwoopt{\citepads}[3][][]{%
   \nonstopmode
   \href{http://adsabs.harvard.edu/abs/#3}%
        {\def\hyper@linkstart##1##2{}%
         \let\hyper@linkend\@empty\citep[#1][#2]{#3}}
   \biblink{#3}{\href{http://adsabs.harvard.edu/abs/#3}{ADS}}
   \errorstopmode}            
 \newcommandtwoopt{\citetads}[3][][]{%
   \nonstopmode
   \href{http://adsabs.harvard.edu/abs/#3}
        {\def\hyper@linkstart##1##2{}%
         \let\hyper@linkend\@empty\citet[#1][#2]{#3}}
   \biblink{#3}{\href{http://adsabs.harvard.edu/abs/#3}{ADS}}%
   \errorstopmode}            
 \newcommandtwoopt{\citeyearads}[3][][]{%
   \nonstopmode
   \href{http://adsabs.harvard.edu/abs/#3}%
        {\def\hyper@linkstart##1##2{}%
         \let\hyper@linkend\@empty\citeyear[#1][#2]{#3}}
   \biblink{#3}{\href{http://adsabs.harvard.edu/abs/#3}{ADS}}%
   \errorstopmode}            
\makeatother
\makeatletter
\newcommand{\bibnote}[2]{\@namedef{#1note}{#2}}
\newcommand{\biblink}[2]{\@namedef{#1link}{#2}}
\makeatother

\newcommand{\be}{\begin{equation}}
\newcommand{\ee}{\end{equation}}
\def\kms{km\,s$^{-1}$}

\def\h2{\ensuremath{\mathrm{H}_2}}

\def\CaII{\ion{Ca}{2}}
\def\Cair{\CaII\ 854.2~nm}
\def\RH{{RH}}
\def\kms{\mbox{{km~s$^{-1}$}}}

\def\MgII{\ion{Mg}{2}}

\def\FeI{\ion{Fe}{1}}
\def\hk{{h\&k}}
\def\MgIIhk{\mbox{\MgII\, \hk}}

\def\RH{{\it RH}}

\def\logtau{\ensuremath{\log_{10} \tau_{500}}}

\def\ackno{{\footnotesize
We thank Bart De Pontieu for his comments on the manuscript.  {We are most grateful to Han Uitenbroek for developing RH.}
This study has been supported by grants from the Swedish Research Council (2015-03994), the Swedish National Space Board (128/15), 
and the Knut and Alice Wallenberg Foundation (\textsc{chromobs}).
AAR acknowledges financial support from the Spanish Ministry of Economy and Competitiveness (AYA2014-60476-P) and the Ram\'on y Cajal fellowships.
We used resources provided by the Swedish National Infrastructure for Computing (SNIC) at the National Supercomputer Centre in Sweden with project-id snic2016-1-131.
IRIS is a NASA small explorer mission developed and operated by LMSAL with mission operations executed at NASA Ames Research center and major contributions to downlink communications funded by ESA and the Norwegian Space Centre. 
This study was discussed in meetings of the group
\emph{Heating of the Magnetic Chromosphere} at the International Space
Science Institute (ISSI) in Switzerland. 
}}

\begin{document}

\title{Non-LTE inversions of the MG II H\&K and UV triplet lines}

\author{Jaime de la Cruz Rodr\'iguez$^{1}$}
\author{Jorrit Leenaarts$^{1}$}
\author{Andr\'es Asensio Ramos$^{2,3}$}

\affil{$^1$ Institute for Solar Physics, Dept. of Astronomy, Stockholm University,
AlbaNova University Centre, SE-106 91 Stockholm Sweden}
\affil{$^2$  Instituto de Astrof\'{\i}sica de Canarias, 38205, La Laguna, Tenerife, Spain}
\affil{$^3$  Departamento de Astrof\'{\i}sica, Universidad de La Laguna, E-38205 La Laguna, Tenerife, Spain}

 \date{Received; accepted}

\begin{abstract}
  The \MgIIhk\ lines are powerful diagnostics for studying the solar chromosphere. They have become particularly popular with the launch of the IRIS 
  satellite, and a number of studies that include these lines have lead to great progress in understanding chromospheric heating, in many 
  cases thanks to the support from 3D MHD simulations.
In this study we utilize another approach to analyze observations: non-LTE \emph{inversions} of the \MgIIhk\ and UV triplet lines including the effects of partial redistribution. Our inversion code attempts to construct a model atmosphere that is compatible with the observed spectra. We have assessed the capabilities and limitations of the inversions using the FALC atmosphere and a snapshot from a 3D radiation-MHD simulation. We find that \MgIIhk\ allow reconstructing a model atmosphere from the middle photosphere to the transition region. We have also explored the capabilities of a multi-line/multi-atom setup, including the \MgIIhk, the \Cair\ and the \ion{Fe}{1}~630.25 lines to recover the full stratification of physical parameters, including the magnetic field vector, from the photosphere to the chromosphere. Finally, we present the first inversions of observed IRIS spectra from quiet-Sun, plage and sunspot, with very promising results.
\end{abstract}
          
  
\section{Introduction} \label{sec:introduction}

The solar chromosphere forms the  {interface region} between the solar interior and the hot corona. Understanding the chromosphere is of great interest to solar and stellar physics because the energy that heats the corona, the mass lost in the solar wind, and the magnetic flux that forms the coronal magnetic field pass through it.
It is difficult, though, to derive the physical state and evolution of the chromosphere because it shows complex spatial and temporal structuring down to scales that cannot be resolved with current  {instruments}. In addition, the UV and optical continua and lines that make up the bulk of the diagnostics form under optically-thick non-LTE conditions so that a straightforward interpretation of the observed spectra is not possible.
A promising avenue into systematically interpreting spatially-resolved chromospheric observations are non-LTE inversions. Such inversions create a best-fit atmospheric model that is consistent with the observations 
(e.g., \citeads{2000ApJ...530..977S}; \citeads{2015A&A...577A...7S}).

The Interface Region Imaging Spectrograph 
\citepads{2014SoPh..289.2733D}
has provided a new spatially and temporally resolved view of the upper chromosphere. The richest IRIS diagnostics are the 
\MgIIhk\ and \MgII~UV triplet lines
(\citeads{2013ApJ...772...90L}; \citeads{2015ApJ...806...14P}).
We present a new inversion code that can take into account non-LTE lines and continua including the effects of partial redistribution for multiple atoms simultaneously. We apply the code to synthetic spectra as well as IRIS observations. 

The broad wings of the \MgIIhk~lines, and the finite spectral window observed with IRIS lead to a limited sensitivity of the lines to photospheric conditions. In addition, the calculations by, e.g., 
\citetads{1957ApJ...126..318G} and \citetads{2015ApJ...809L..30C}
indicate a level of degeneracy in the response of the intensity profile with respect to variations in temperature and microturbulence. We therefore also investigate whether multi-line, multi-atom inversions using a combination of the \MgIIhk, \MgII~UV triplet, \Cair\ and \FeI~630.2~nm yield improved inference of the atmosphere,  {among others that could also be used}.

\section{Method} \label{sec:method}
\begin{figure*}
  \includegraphics[width=\textwidth]{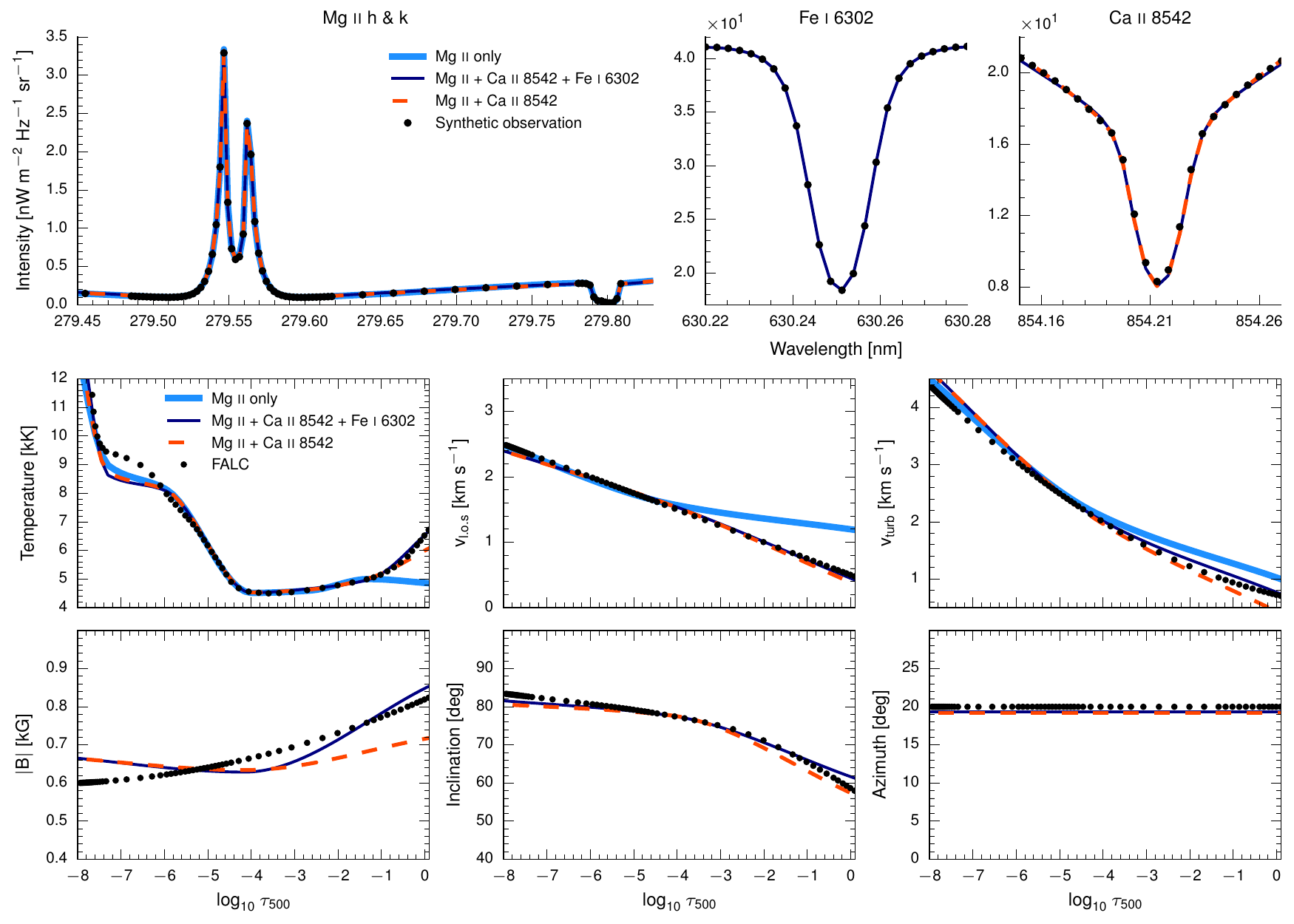}
\caption{Inversions of a modified FALC 1D model atmosphere. Black dots indicate the input model and the resulting synthetic observations. The blue curve is an inversion using the \MgIIhk\ profiles only, the black curve using both \MgIIhk\ and \Cair, and the red dashed curve using \MgIIhk, \Cair, and \FeI\ 630.2~nm. The upper row displays the synthetic observations and profile fits, while the lower two rows show the input atmosphere model and the inferred atmospheres.
\label{fig:FALC}}
 \end{figure*}
We have incorporated a non-LTE radiative transfer engine based on the \RH\ code
(\citeads{2001ApJ...557..389U})  {to a newly developed inversion framework including
some of the ideas presented in}
%
%
\citetads{2015A&A...577A.140A}.
%

\RH\ has become the workhorse non-LTE radiative transfer code in solar physics owing to its flexible nature and comprehensive inclusion of the relevant physics. An important aspect of \RH\ is that it includes the effect of partial redistribution (PRD), including a fast approximation of angle-dependent PRD
(\citeads{2012A&A...543A.109L})
that allows for fast computation of accurate \MgIIhk\ line profiles in models with varying vertical velocity.

 In our inversion code we run \RH\ to solve the non-LTE radiative transfer problem for a given stratification of temperature, hydrogen populations, electron density, vertical velocity and microturbulence as a function of 
optical depth. Response functions are computed numerically.
The model atmosphere is initialized by specifying the temperature, velocity and microturbulence for a given optical-depth scale. The hydrogen density, electron density and height scale are then computed using an LTE equation of state
(\citeads{2016arXiv160606073P})
 and hydrostatic equilibrium.  {The inversions can be performed in full-Stokes mode or intensity-only mode. When mixing spectral region with and without polarization, the code runs in full-Stokes mode and the unpolarized regions are given zero weight in Stokes~$Q$, $U$ and $V$.}
 
The physical quantities of the model are discretized using a small number of node points
 (\citeads{1992ApJ...398..375R}).
To produce a fully depth-stratified atmosphere the nodes are connected using a second-order piece-wise Bezier-spline 
(\citeads{2013ApJ...764...33D}).
The depth resolution is given by the number and location of the nodes that are used for each physical parameter.

We assume plane-parallel geometry  {and statistical equilibrium} to calculate the atomic level populations in each pixel  {(1.5D)}, {a necessary compromise} to make the inversion problem treatable. We compute the response function based on the wavelength grid and instrumental spectral smearing of the near-UV window of the IRIS spectrograph ($\delta\lambda=2.54$~pm). To speed up the calculations we have used a coarser sampling of 20.36~pm in the extended wings of the \MgIIhk\ lines but we kept the full sampling close to the line cores. This reduced the number of wavelengths by a factor three.  {The result of the inversion is insensitive to very fine sampling in the broad and slow-varying wings of \MgIIhk.}
%
%

For the multi-line, multi-atom inversions we include a frequency grid based on critically sampled line profiles with the CRISP instrument
on the Swedish 1-m Solar telescope
(\citeads{2003SPIE.4853..341S}; \citeads{2008ApJ...689L..69S}).

 {In \S\ref{sec:falc} and \S\ref{sec:bif} we have computed synthetic observations with RH working in full-Stokes mode and assuming 1.5D radiative transfer. We used the same fast PRD approximation that is used during the inversions.}
%
\section{Results} \label{sec:results}

\subsection{Inversions of the FALC model}\label{sec:falc}
\begin{figure*}
  \includegraphics[width=\textwidth]{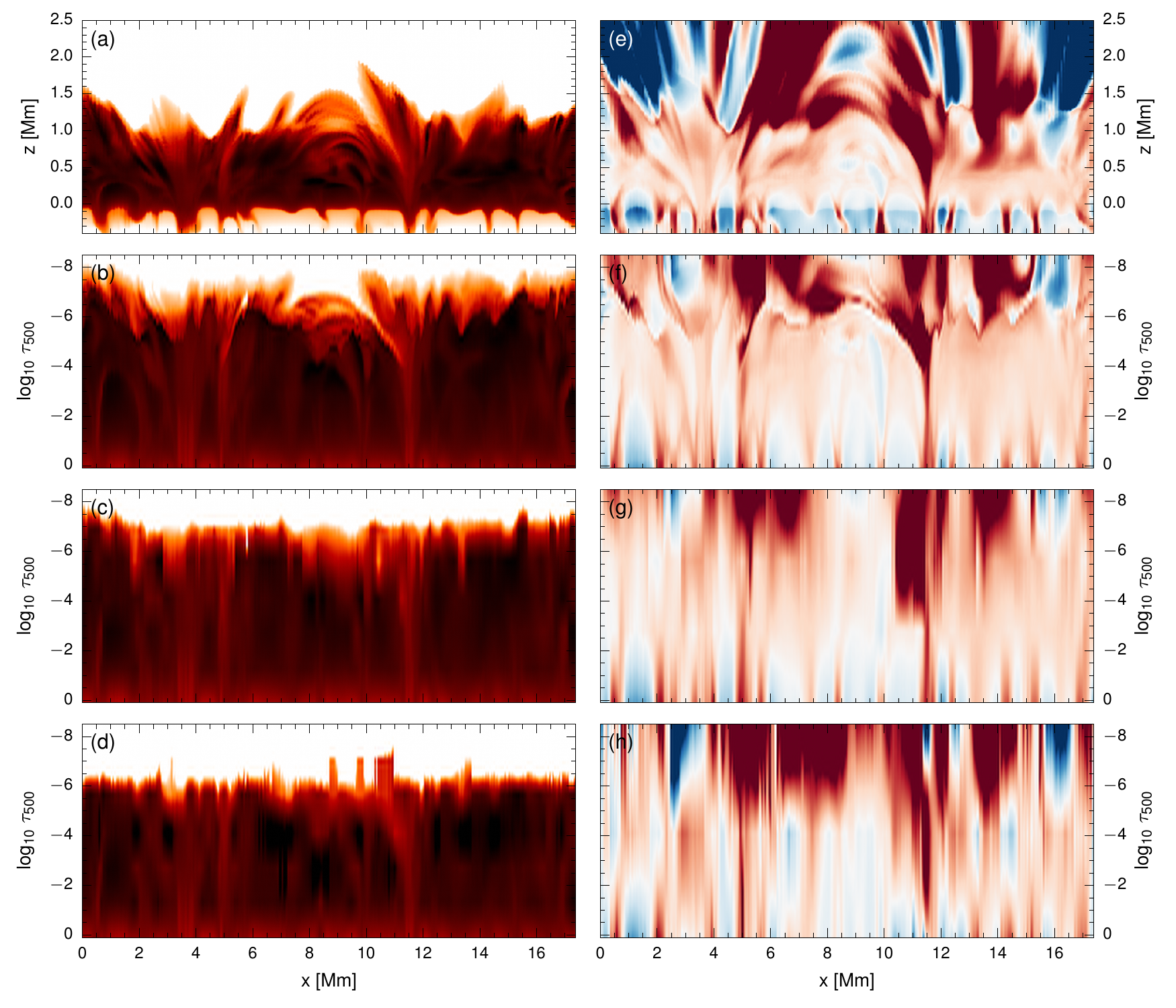}
\caption{Vertical slice through the simulation snapshot. \emph{Left-hand panels}: Temperature. \emph{right-hand panels}: line-of-sight velocity. \emph{Top to bottom}: the original model, the original model on a regular $\logtau$ depth-scale, the original model degraded to a representation using seven nodes, and the results from the inversions.
\label{fig:bif}}
 \end{figure*}
 
\begin{figure*}
  \includegraphics[width=\textwidth]{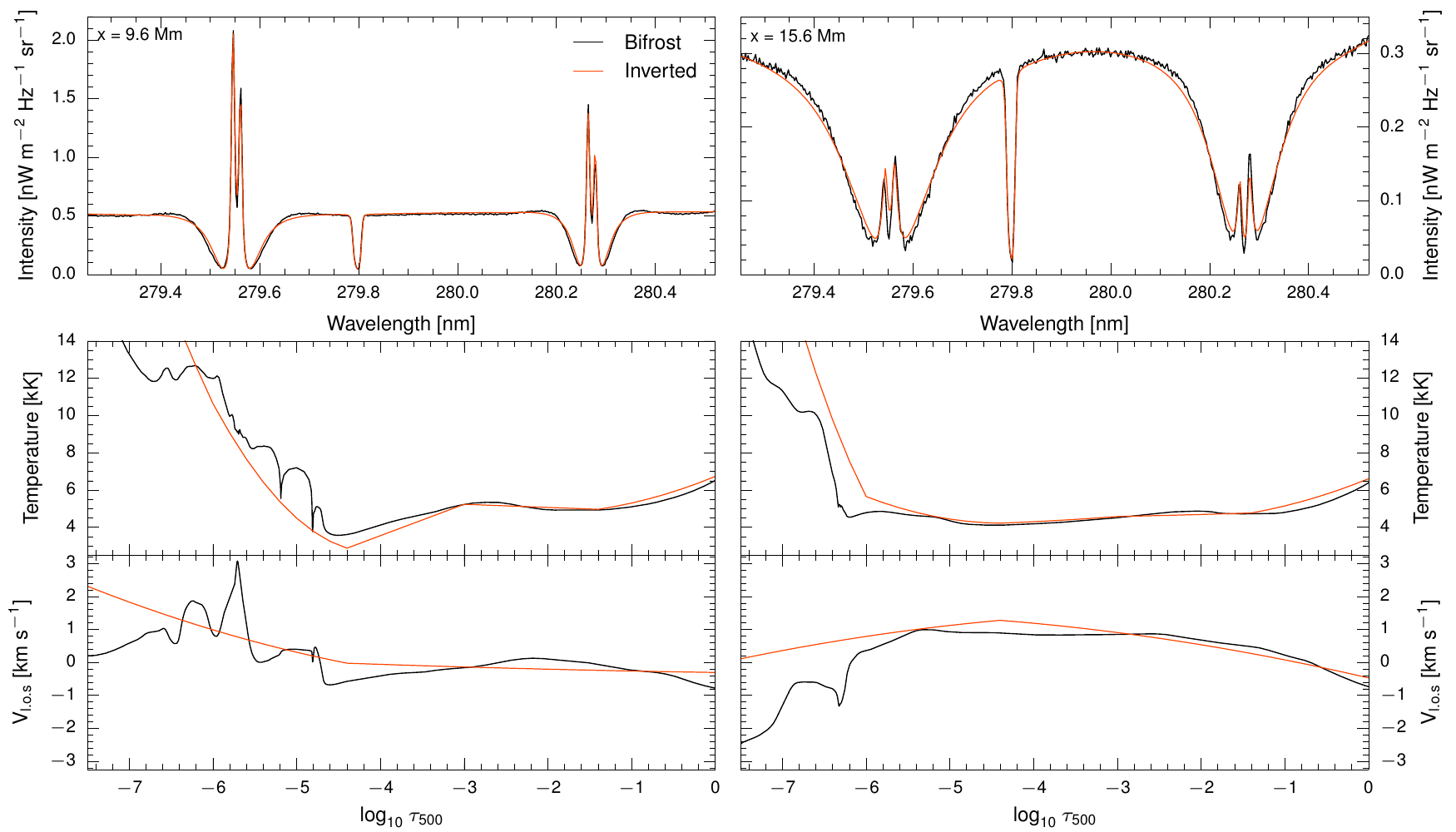}
\caption{Inversion of two \emph{Bifrost} columns. The input (black) and inverted (red) spectra are shown in 
the top panels. The middle and bottom rows show the stratification of temperature (middle) and vertical velocity (bottom) for the input and inverted models with the same color coding.
\label{fig:profs}}
 \end{figure*}

As a first test we inverted line profiles from the FALC~1D semi-empirical model atmosphere of 
\citet{1993ApJ...406..319F}.
We changed the microturbulence from the original form to a smoother slightly convex profile in $\log_{10}~\tau_{500}$, added a magnetic field  {with gradients} in strength and inclination and a constant azimuth, and added a linear gradient in vertical velocity  {as a function of optical-depth}. The inversions were performed in two cycles, improving the convergence and speed of the inversion 
(\citeads{2007ApJ...670..885P}).
In the first cycle we used four nodes in temperature, three nodes in vertical velocity, two nodes in microturbulence and one node in magnetic field strength, azimuth and inclination. 

The second cycle was initialized from the atmosphere inferred in the first cycle using seven nodes in temperature, three nodes in vertical velocity and microturbulence, magnetic field strength and inclination, and one node in azimuth. We have given zero weight to Stokes $Q$, $U$ \& $V$ in the \ion{Mg}{2} spectral window to simulate an IRIS observation. 

Figure~\ref{fig:FALC} shows the results. The inversions succeed in matching the synthetic observations precisely. The temperature is matched well for $-5< \logtau<-1.5$ in all cases. 
 {In the upper chromosphere (\logtau$=[-7.3,-6]$) the inversions are worse since the radiation is only weakly coupled to the local conditions of the model at those heights.}
Nevertheless the code manages to mimic the shape of the transition region relatively well. According to this test, the \MgII\ lines alone (light-blue) can constrain the temperature stratification reasonably well. The reconstruction of the deeper photosphere improves when the \Cair\ line is included (orange-red), and it is almost perfect when we include the \ion{Fe}{1}~line (navy-blue). 
 \begin{figure*}
  \includegraphics[width=\textwidth]{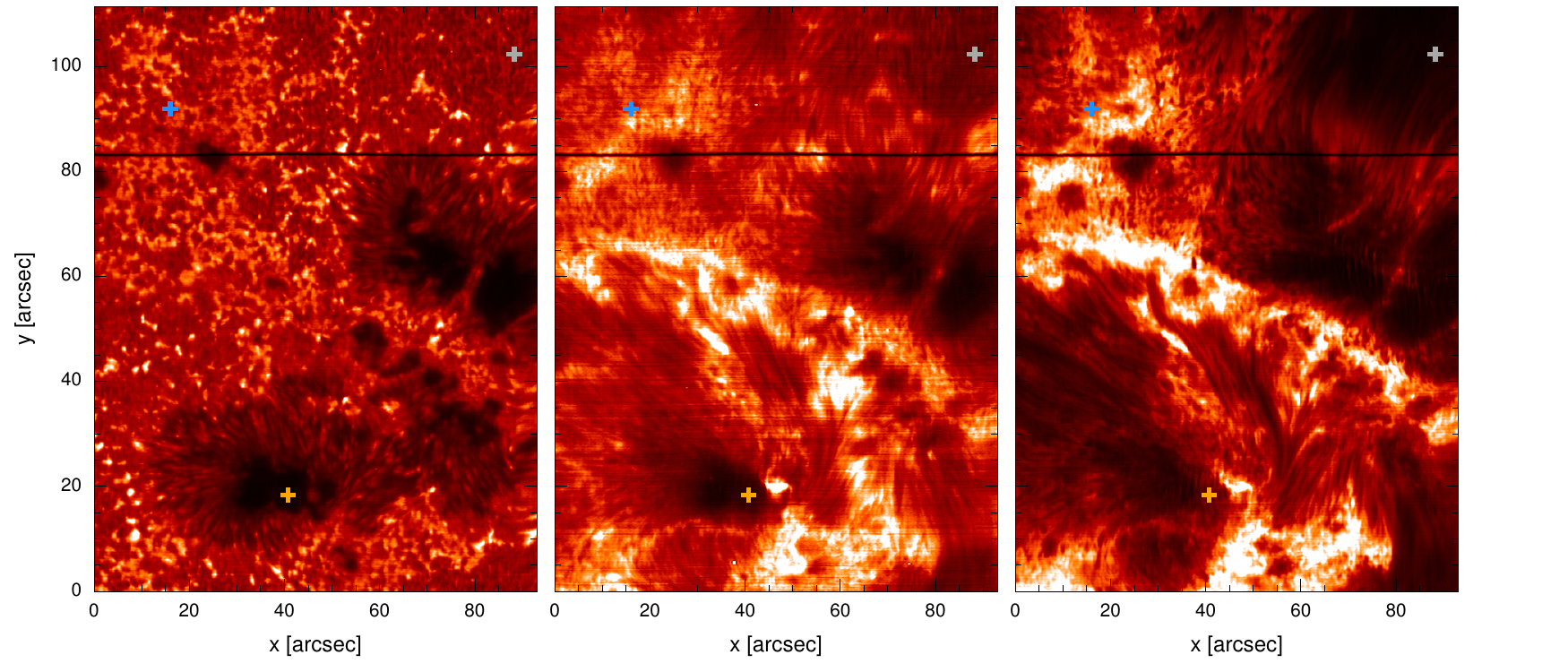}
  \includegraphics[width=\textwidth]{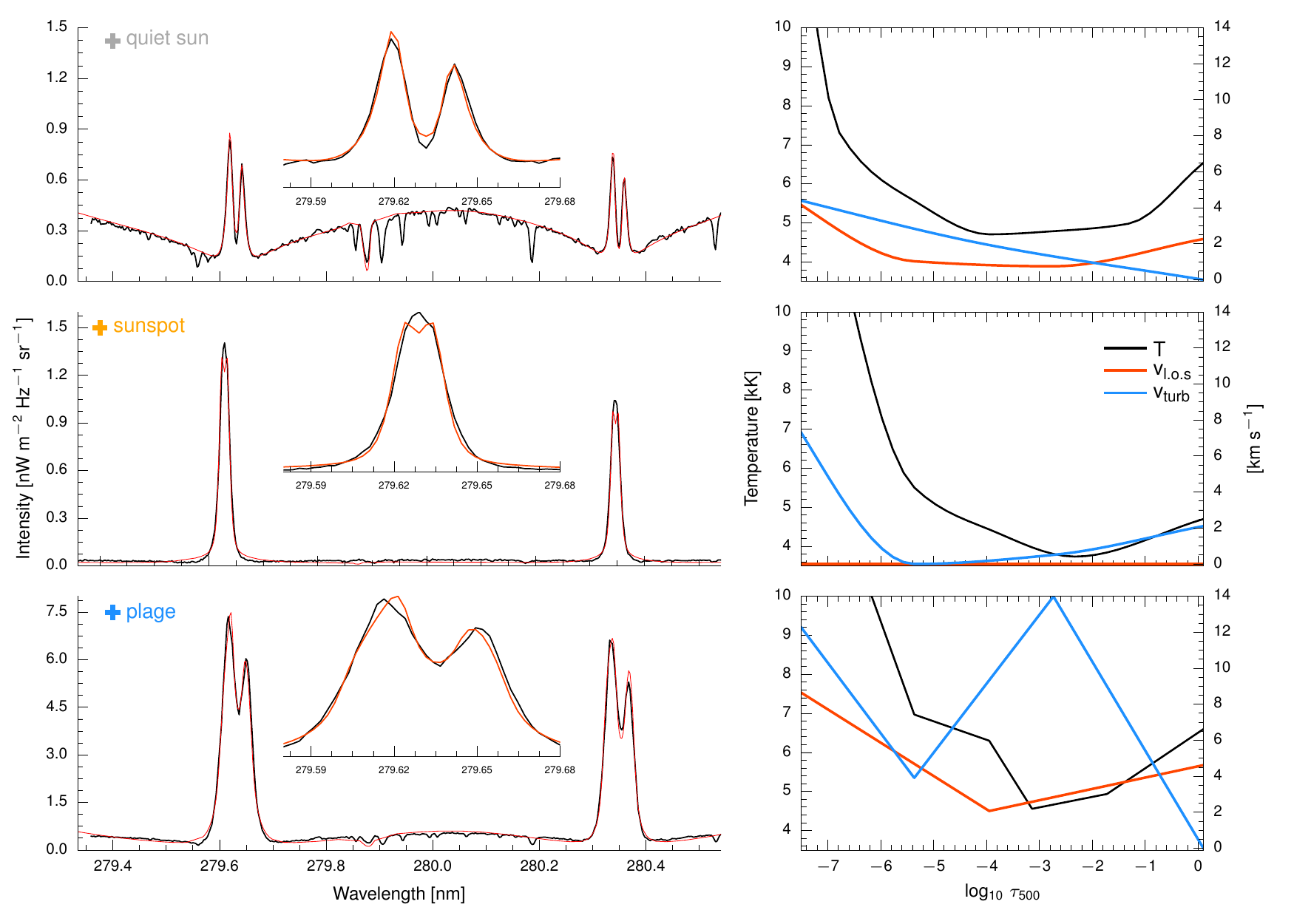}
\caption{\emph{Top row:} Monochromatic images illustrating the location of our selected pixels for the inversion taken in the blue wing of the k line, in the core of the UV line and in the core the k line. \emph{Left column:} Observed spectra (black) and best-fit (red) for quiet-Sun, sunspot and plage pixels respectively from top to bottom. The inset depicts a zoom in the core of \ion{Mg}{2}~k. The reconstructed temperature (black), $v_{l.o.s}$ (red) and $v_{turb}$ (blue) stratifications are shown in the right column.
\label{fig:iris}}
 \end{figure*}
The vertical velocity and microturbulence stratifications are retrieved by the inversion scheme relatively well at all depths, but adding the \ion{Fe}{1} and the \ion{Ca}{2} lines improves the estimate of the vertical velocity in the deepest layers of the atmosphere. These results are somewhat expected because the broad wings of the \MgII\ lines are sensitive to the temperature stratification, but the small intensity gradient makes the wings insensitive to changes in the vertical velocity at $\logtau>-3$. 

The \ion{Ca}{2} line constrains the magnetic field strength between $-6<\logtau<-4$, but the shape of the gradients are not reproduced very well. Only when the \ion{Fe}{1} line is included the entire field strength profiles are properly retrieved for this configuration of the magnetic field. The {magnetic} field inclination is reconstructed accurately in both cases for the assumed noise levels. All combined, these lines {appear} to be a powerful tool to retrieve the full depth-stratification of the solar atmosphere, including the magnetic field vector. 

This test also shows that even in a quiet-Sun model as FALC there is enough coupling to the local physical conditions present in the input model for the inversion to work. The FALC test model atmosphere is very smooth, so in \S\ref{sec:bif} we test inversions of a 3D~MHD simulation containing strong gradients.


\subsection{Inversions of a radiation-MHD model}\label{sec:bif}
{In this section we present a setup that can be used to invert IRIS~\ion{Mg}{2}~data alone, as co-observations with other facilities are not always available.} 

We have calculated synthetic profiles from a vertical slice through from {the publicly available} 3D~MHD simulation of \emph{enhanced network} that was performed with the \emph{Bifrost} code 
(\citeads{2011A&A...531A.154G}; \citeads{2016A&A...585A...4C}, {snapshot~385}).
This slice cuts through two patches of magnetic bright points and through the fibril-like features that are formed between them, {extending from pixel} $(x,y)=(69,206)$ to $(x,y) = (431,339)$. This simulation includes the effects of non-equilibrium hydrogen ionization that has an effect on the electron density in the chromosphere and transition region 
(\citeads{2007A&A...473..625L})
and unlike FALC, it is not in hydrostatic equilibrium. We have degraded the synthetic profiles to the spectral resolution of IRIS and have added Gaussian noise with standard deviation
$\sigma=2\times10^{-3}$. We have estimated a similar noise level from the wings of the \ion{Mg}{2}~lines in IRIS observations.

Our experiment with the FALC model {provides} evidence that velocities in the photosphere are poorly constrained when the \ion{Mg}{2} lines are {used exclusively}. We have also included the \ion{Ni}{1}~281.4350~nm line to better constrain velocities in the photosphere. This is one of the weakest, relatively unblended, lines in the same IRIS spectral window as \MgIIhk. Because the line is weak we can include it in the inversion assuming LTE. We suspect there are more similar lines that could in principle also be included. 

We ran the inversions in two cycles as described in Section~\ref{sec:falc}, ending with 7 nodes in temperature and 4 nodes in vertical velocity. Our results are presented in Figure~\ref{fig:bif}. The top row displays the temperature and vertical velocity on the original height scale of the simulation. The second row shows these properties on a per-pixel $\logtau$ scale. The third row displays how the slices look if we represent them using 7 nodes in temperature and 4 nodes in vertical velocity connected by Bezier-splines. This illustrates the kind of depth resolution that we can expect from the inversions. Finally, the inversion results are shown in the last row.

{Temperature} \ The inverted temperatures shown in (d) resemble those in (b), given the limitations of the node representation. In fact, there are regions that resemble the original atmosphere better than those in the interpolated model in (c), probably because the inversion can adjust the values of the nodes to fit the gradient of the spline that is connecting them, whereas the plain interpolation in (c) is very sensitive to the local value of the atmosphere {at the locations of the nodes}. Close to the transition region, around $\logtau=-6$, there are atmospheric features that are well reconstructed {in the majority of the columns}, but the overall transition region temperature increase appears at larger optical depth in the inversion. 

{Line-of-sight velocity} \ The reconstructed line-of-sight velocities are shown in panel (h) of Figure~\ref{fig:bif}. In comparison with the original bifrost slice in (f), the resemblance is a bit worse {than in the temperature case}. We have experimented with the number of nodes used in this parameter with similar results: velocities in the very upper layers of the model typically have the same sign as the original slice but larger amplitudes. 

{Figure~\ref{fig:profs} shows in detail the results from the inversion of two \emph{Bifrost} columns ($x=9.6$ and $x=15.6$~Mm).
These plots illustrates that our code can reproduce the large scales present in the input model atmosphere, but not the small scale features. The node-based parameterization is not able to reproduce sharp plato-like areas in the upper chromosphere, like the one in the left column around \logtau$=-6$. This is probably one of the reasons why we get a hotter transition region in some columns. However, the effect of not including non-equilibrium hydrogen ionization can also be partly responsible in some areas. Our EOS assumes LTE ionization and the inversion can compensate the lower LTE electron density by increasing the temperature close to the transition region.}


\subsection{Inversions of IRIS data}\label{sec:IRIS}

We {applied} our inversion code with an IRIS observation of active region NOAA~AR~12104 acquired on 04-Jul-2014 at 11:40~UT. The dataset consists of a 400 step raster with an exposure time of 30~s per slit position and 5.3~pm spectral resolution. This observation was previously used by 
\citetads{2015ApJ...809L..30C}.

We inverted three spectra formed in quiet-Sun {(grey-cross)}, plage {(blue-cross)} and quiescent sunspot {(orange-cross)} regions. The field-of-view and the location of the pixels are shown in Figure~\ref{fig:iris}. The images in the top row depict monochromatic images in the blue wing of the k line, in the core of one of the subordinate triplet lines and in the core of \MgII~k. 


\begin{figure*}
  \includegraphics[width=\textwidth]{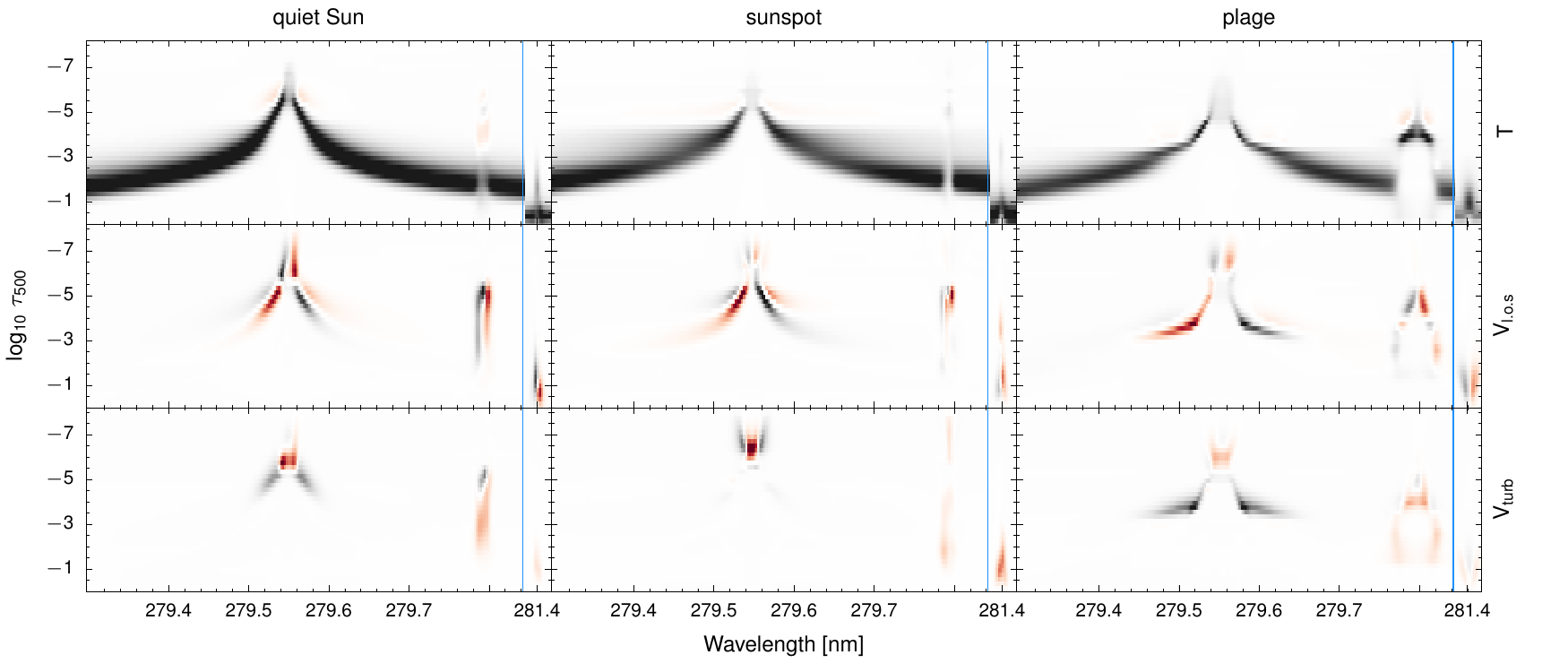}
\caption{From left to right: Response functions for the quiet-Sun, sunspot and plage inverted models. From top to bottom: temperature, velocity and microturbulence. For a better visualization, we have normalized the response function at each wavelength by the corresponding emerging intensity. The areas in black have a positive response to a positive perturbation of the physical parameters, whereas red indicates a decrease of the intensity.
\label{fig:rf}}
 \end{figure*}
We included the photospheric \ion{Ni}{1}~line that we used in the previous experiment. The inversions were again performed in two cycles, ending with 7 nodes in temperature in all cases. We used 4 nodes in line-of-sight velocity and 3 nodes in microturbulence to invert the plage and quiet-Sun profiles. The sunspot profile was inverted using one node in velocity (see discussion below). We initialized the inversions with a smoothed version of FALC with a constant microturbulence of 3~\kms. We chose the umbra of the sunspot during quiescence as a rough zero velocity reference.

Our results are shown at the bottom of Figure~\ref{fig:iris}. The code fits all three cases quite well.
The best fit to the quiet-Sun profile is very good except for small discrepancies in the core of some of the lines. The resulting model has a cooler chromosphere and transition region than FALC, which is somewhat expected given that the h$_2$ and k$_2$ intensities are lower than those predicted from FALC. The inversion code has reconstructed a velocity gradient needed to reproduce the location and the asymmetries close to line center. The l.o.s velocity in the photosphere is mostly constrained by the \ion{Ni}{1}~line. The microturbulence profile starts at zero in the photosphere and it increases almost linearly to 3 \kms\ in the upper chromosphere, which is similar to the average chromospheric value of the FALC model.
 
We have fitted the quiescent sunspot profile with a single node in velocity. Adding more nodes did not improve the best-fit and we found out that including more nodes could also lead to a solution where the central line depressions are suppressed by introducing a very step gradient in velocity, instead of filling up the core with adjustments in the temperature profile. The observed profile is also very well fitted, except for the innermost line positions in the h\&k lines. Further experiments with the number of nodes showed that the inner core is sensitive to the exact location of the transition region. 
{The reconstructed temperature profile reaches a minimum  minimum at $\log_{10}\tau_{500}=-2.5$ and it smoothly increases towards the chromosphere. The transition region is located at higher optical-depth than in the quiet-Sun example.} The code sets the microturbulence to almost zero except for the very upper layers where it increases again.
 
The plage profile is qualitatively well reproduced by our best fit. We have increased the gas pressure at the upper boundary of the model atmosphere by more than one order of magnitude (like \citeads{1974SoPh...39...49S}), which is used to start the hydrostatic equilibrium integration. Otherwise we could not get sufficient emission in  the line cores even for very high temperatures. 
The temperature stratification that we derive is surprisingly similar to the model presented in
 \citetads{2015ApJ...809L..30C},
 which is reassuring because we did not impose any particular shape on the solution: a steep increase in temperature at $\logtau=-3$, followed by an almost flat chromosphere and then a {steep} transition region.

The inversion code struggled to fit the plage profile when we used 3 nodes in microturbulence, particularly around the k$_1$ and h$_1$ minima. With 4 nodes in microturbulence the code introduces very high values around $\logtau=-3$. Therefore most of this microturbulence is affecting the spectra around  h$_1$ and k$_1$. Overall, the derived values of the microturbulence are surprisingly high ($\sim8$~\kms\ close to $\logtau=-4.0$), which are consistent with previous studies (\citeads{1974SoPh...39...49S}; \citeads{2015ApJ...799L..12D}; \citeads{2015ApJ...809L..30C}). In this case we had to use straight lines instead of a Bezier curve to connect the nodes because it allows to define steep gradients in a much better way.
 
In Figure~\ref{fig:rf} we show response functions (RFs) computed from the derived quiet-Sun, sunspot and plage inversions. RFs illustrate how the spectra react to small perturbations in the physical parameters at each depth point. The rows correspond to temperature, line-of-sight velocity and microturbulence from top to bottom. The broad wings of the \ion{Mg}{2}~h\&k~lines are sensitive to the middle/upper photosphere, whereas the core samples almost up to $\logtau = -7.5$. The sensitivity to line-of-sight velocities and to microturbulence is (as expected) particularly strong in the inner part of the lines and very weak in the extended broad wings. {The response function of the \ion{Ni}{1}~line peaks in the photosphere, explaining why its addition is a good complement to the \ion{Mg}{2}~lines.}


In this section we have presented the first \ion{Mg}{2} h\&k inversion from three chromospheric regimes in the solar atmosphere. The sensitivity of the \ion{Mg}{2}~h\&k lines to the location of the transition region sets strong constraints on the inversion of temperatures and sometimes velocities. 
\section{Discussion and conclusions} \label{sec:conclusions}

We have included the effect of PRD in a non-LTE inversion code to study the \ion{Mg}{2}~h\&k lines. We have performed tests with the FALC model atmosphere and with a 3D MHD simulation to show that the inversion code is able to retrieve information in the upper layers of the solar chromosphere.

Throughout this study we have assumed that we can model the \ion{Mg}{2}~h\&k lines using plane-parallel radiative transfer, which can be a serious approximation in the core of strong lines. The 1.5D approximation used in our inversions usually leads to higher contrast in the emerging intensity maps when compared to a full 3D evaluation because horizontal radiative transfer has a smoothing effect. The inversion code normally compensates for 3D effects by decreasing the contrast in the retrieved temperature maps (\citeads{2012A&A...543A..34D}). The results of 
\citetads{2016arXiv160605180S} 
indicate that the plane-parallel approximation is reasonable for inversions with the \ion{Mg}{2}~h\&k lines at disk center, but one should keep in mind that this approximation is a \emph{varying} source of errors that depend on the structure of the atmosphere surrounding our pixel. Especially close to large horizontal gradients in the atmosphere the errors may be large.

Out results with the radiation-MHD model are particularly encouraging because this model includes the effects of non-equilibrium hydrogen ionization and is not in hydrostatic equilibrium, yet the inversions can in most cases retrieve a similar temperature stratification to that in the input model atmosphere and provide good estimates of line-of-sight velocities even when only the \ion{Mg}{2}~h\&k and UV lines are included. The addition of more diagnostics, especially those sensitive to different regimes and temperatures in the atmosphere should help to further constrain the quality of the reconstruction.

We have presented the first inversions of spatially-resolved IRIS spectra from typical quiet-Sun, plage and sunspot profiles. Our plage inversions require very high values of microturbulence in the upper photosphere to reproduce the width of the profile, which in itself is very interesting because it points to a physically different situation compared to quiet-Sun and sunspots. Unfortunately we do still not  understand the nature of this enhanced microturbulence. We believe that the addition of more non-LTE diagnostics from other atomic species can help to understand it. The \ion{Ca}{2}~H\&K lines and the \ion{Ca}{2}~IR~triplet lines are very good candidates {for such an addition.}

\begin{acknowledgements}
\ackno
\end{acknowledgements}

\bibliographystyle{aa} 
\bibliography{rhinv}

\end{document}